\documentclass[aps,prb,twocolumn,groupedaddress]{revtex4}
\usepackage{graphicx}

\begin{document}
\title{Effects of the spin-orbital coupling on the vacancy-induced magnetism on the honeycomb lattice}
\author{Weng-Hang Leong}
\author{Shun-Li Yu}
\author{Jian-Xin Li}
\affiliation{ National Laboratory of Solid State Microstructure
and Department of Physics, Nanjing University, Nanjing 210093,
China}
\date{\today}

\begin{abstract}
The local magnetism induced by vacancies in the presence of the
spin-orbital interaction is investigated based on the half-filled
Kane-Mele-Hubbard model on the honeycomb lattice. Using the self-consistent
mean-field theory, we find that the spin-orbital coupling will enhance the
localization of the spin moments near a single
vacancy. We further study the magnetic structures along the zigzag edges formed by a chain of vacancies. We find that the spin-orbital coupling tends to suppress the counter-polarized ferrimagnetic order on the upper and lower edges, because of the open of the spin-orbital gap. As a result, in the case of the balance
number of sublattices, it will suppress completely this kind of ferrimagnetic order. But, for the imbalance case, a ferrimagnetic order along both edges exists
because additional zero modes will not be affected by the spin-orbital coupling.
\end{abstract}
\maketitle

\section{INTRODUCTION}

Graphene and related nanostructured materials have attracted much
interest in solid state physics recently due to their
bidimensional character and a host of peculiar
properties~\cite{Neto}. Among them, the investigation of the magnetic properties
in graphene is one of the fascinating topics, as no $d$ and $f$ elements are necessary in the
induction of magnetism in comparison with the usual magnetic
materials. Theoretical predictions and experimental investigations
have revealed that a nonmagnetic defect such as an impurity or
a vacancy can induce the non-trivial localized
magnetism~\cite{Yazyev,Hirashima,Yazyev2,Ugeda,Nair}. Similarly, a
random arrangement of a large number of vacancies which are generated by the high-dose exposure of graphene to strong electron
irradiation~\cite{Meyer} can also induce magnetism
theoretically~\cite{Fern2}. These studies not only have
the fundamental importance, but also open a door for the possibility of application in new technologies for
designing nanoscale magnetic and spin electronic devices.

On the other hand, the topological insulating electronic phases driven
by the spin-orbital (SO) interaction have also attracted much
interest recently. The Kane-Mele model for the topological band
insulator is defined on the honeycomb lattice~\cite{Kane,Kane2}
which is the same lattice structure as graphene. Possible
realization of an appreciable SO coupling in the honeycomb lattice
includes the cold fermionic atoms trapped in an extraordinary
optical lattice~\cite{Lee}, the transition-metal oxide
Na$_{2}$IrO$_{3}$~\cite{Shitade} and the ternaty compounds such as
LiAuSe and KHgSb~\cite{Hai}. Topological band insulator has a
nontrivial topological order and exhibits a bulk energy gap with
gapless, helical states at the edge~\cite{Zhang,Konig,Hasan}.
These edge states are protected by the time reversal symmetry and are
robust with respect to the time-reversal symmetric perturbations, such
as non-magnetic impurities. It is shown that a vacancy, acting as a minimal
circular inner edge, will induce novel time-reversal invariant bound states
in the band gap of the topological insulator~\cite{Shen,Gonz,Liang}. Theoretically,
it is also shown that the SO coupling suppresses the edge magnetism
induced in the zigzag ribbon of the honeycomb lattice in the presence of
electron-electron interactions~\cite{Fern}. Thus, it is expected that
the SO coupling would also affect the local magnetism in the bulk induced by
vacancies.

In this paper, we study theoretically the effects of the SO coupling
on the local magnetism induced by a single and a multi-site vacancy on the honeycomb lattice,
based on the Kane-Mele-Hubbard model where both the SO coupling and the Hubbard interaction between electrons
are taken into consideration. This model has been extensively studied to explore the effect of the strong correlation on the topological insulators \cite{SRachel,MHohenadler,SLYu,DZheng,WWu,CGriset,JWen}.
Making use of the self-consistent mean field approximation, we calculate the
local spin moments and their distribution around the vacancies. For a single vacancy, we find that the main effect of the SO coupling is to localize the spin moments to be
near the vacancy, so that it will enhance the local spin moments. For a large stripe vacancy by taking out a chain of sites from the lattice, we find that the SO coupling
tends to suppress the counter-polarized ferrimagnetic order induced along the zigzag edges, because of the open of the SO gap. As a result, in the case of the balance
number of sublattices (with even number of vacancies), the SO coupling will suppress completely the counter-polarized ferrimagnetic order along the upper and lower edges.
While, in the case of the imbalance number of sublattices (with odd number of vacancies), a ferrimagnetic order along both edges exists
because additional zero modes will not be affected by the SO coupling.

We will introduce the model and the method of the self-consistent mean-field approximation in Sec.II. In Sec.III and IV, we present the results for a single vacancy and a multi-site vacancy, respectively.
Finally, a brief summary will be given in Sec.V.

\section{MODELS and COMPUTATIONAL METHODS}

We start from the Kane-Mele model~\cite{Kane}, in which the
intrinsic SO coupling with a coupling constant $\lambda$ is
included.
\begin{equation}
H_0=-t\sum_{\langle ij\rangle,\sigma}c^\dag_{i\sigma}
c_{j\sigma}+i\lambda\sum_{\langle\langle ij\rangle\rangle\sigma\sigma'}v_{ij}\sigma^z_{\sigma\sigma'}
c^\dag_{i\sigma}c_{j\sigma'},
\end{equation}
where $c^{\dag}_{i\sigma}$ ($c_{j\sigma}$) is the
creation(annihilation) operator of the electron with spin
$\sigma$ on the lattice site $i$, $\langle ij\rangle$ represents
the pairs of the nearest neighbor sites (the
hopping is $t$) and $\langle\langle ij\rangle\rangle$ those of
the next-nearest neighbors. $v_{ij}=+1(-1)$ if the electron makes
a left(right) turn to get to the second bond. The size of our
system is considered to be finite with periodic boundary
condition. So, the position of each lattice site can be described
specifically by $i=\Gamma(m,n)$, representing that the
lattice site $i$ is in the $m$th column and the $n$th row, and
$\Gamma=A,B$ the sublattice labels. The number of the unit
cells is denoted by $N_c=L^2$, therefore the total number of
the lattice sites is $N_l=2L^2$. To
consider the correlation between electrons, we will include the
Hubbard term in the Hamiltonian, which is given by $H_I$,
\begin{equation}
H_I=U\sum_{i}\hat{n}_{i\uparrow}\hat{n}_{i\downarrow},
\end{equation}
where $\hat{n}_{i\sigma}=c^\dag_{i\sigma}c_{i\sigma}$. When
vacancies are introduced, the hoppings between the
vacancy and the nearest neighbors and the on-site interaction on
that vacancy are subtracted from the overall Hamiltonian. Hence
the corresponding number of the lattice sites is $N_l=2L^2-N_v$,
where $N_v$ is the number of vacancies. The total number of
electrons $N_e$ is fixed to be at the half-filling ($N_e=N_l$).
The Hubbard interaction term is treated with the self-consistent mean
field approximation, so that we will obtain an effective
single-particle Hamiltonian where the electrons interact with a
spin-dependent potential,
\begin{equation}
H_I\simeq U\sum_{i,\sigma}
\langle\hat{n}_{i-\sigma}\rangle\hat{n}_{i\sigma}-U\sum_{i}\langle\hat{n}_{i\uparrow}\rangle
\langle\hat{n}_{i\downarrow}\rangle.
\end{equation}
And the overall mean field Hamiltonian $H_{mf}$ is then given by,
\begin{equation}
H_{mf}=U\sum_{i\sigma}\langle\hat{n}_{i-\sigma}\rangle\hat{n}_{i\sigma}+H_0.
\end{equation}
After diagonalizing the Hamiltonian $H_{mf}$, we can determine the occupation number
$\langle\hat{n}_{i-\sigma}\rangle$ at each site with different
spins using the eigenvectors of $H_{mf}$, and this process is
carried out iteratively until a required accuracy is reached. Then
the magnetic moment of each site $m_{i}=\langle\hat{n}_{i\uparrow}-\hat{n}_{i\downarrow}\rangle$
can be calculated.
We note that a collinear magnetic texture is assumed in our system, as used before for the investigations of Kane-Mele-Hubbard model~\cite{Fern,SRachel}. We have checked the results with the non-collinear magnetic texture and found that the collinear magnetic texture is favored.

\section{MAGNETISM WITH ONE VACANCY}

\begin{figure}[pt]
\begin{center}
\includegraphics[width=0.46\textwidth]{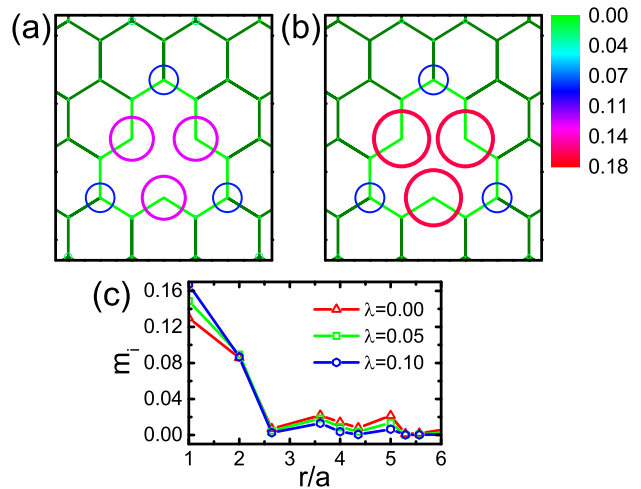}
\caption{(color online). (a) and (b): Distribution of the spin moments $m_{i}$ on
lattice sites around a single vacancy at $A(7,7)$ with $U=1.0t$, in
which (a) corresponds to the SO coupling constant $\lambda=0.0$
and (b)$\lambda=0.1t$. The area and color of the hollow circles
represent the magnitude of the spin moments. (c)$m_{i}$ on the $B$ sublattice as a function of the distance $r$ away from the vacancy. The unit $a$ is the distance between the nearest sites.}\label{loc1}
\end{center}
\end{figure}

The calculation is carried out on the lattice with
$N_c=14\times14$ unit cells in which a single vacancy is
introduced on the site $A(7,7)$. Figure~\ref{loc1} displays the
distribution of the magnetic moment when the Hubbard interaction
is taken to be $U=1.0t$, in which the size and the color of the
circle on each lattice site denote the magnitude of the local
spin moment. From Fig.1(a) where the SO coupling is turned off, one can see
that localized magnetic moments are induced around the vacancy in
the presence of a finite Hubbard interaction $U$. This is in
agreement with the prediction of the Lieb theorem~\cite{Lieb}
regarding the total spin $S$ of the exact ground state of the
Hubbard model on bipartite lattices. It states that the total spin
$S$ is given by the sublattice imbalance $2S=|N_{A}-N_{B}|$, with
$N_{A}$ and $N_{B}$ the number of atoms belonging to each
sublattice. With the introducing of a single vacancy on the $A$
sublattice, an imbalance $N_{B}-N_{A}=1$ appears and a magnetic
structure near the vacancy with the total spin $S=1/2$ will form.
Similar results have also been obtained in recent studies in graphene~\cite{Yazyev,Hirashima,Yazyev2}.

In the presence of the SO coupling, the
magnitude of the magnetic moments around the vacancy increases, as shown in
Fig.~\ref{loc1}(b) for $\lambda=0.1t$. At the meantime, if we
check the distribution of the magnetic moments, as shown in
Fig.1(c) where the magnitude of the magnetic moments on sublattice $B$ as a function of the distance $r$ away from the vacancy is presented,
one will find that the magnetic moments are more localized with the increase of the SO coupling.
These features demonstrate that the SO coupling will enhance
the magnetic moments near the vacancy notably.

\begin{figure}[b]
\begin{center}
\includegraphics[width=0.46\textwidth]{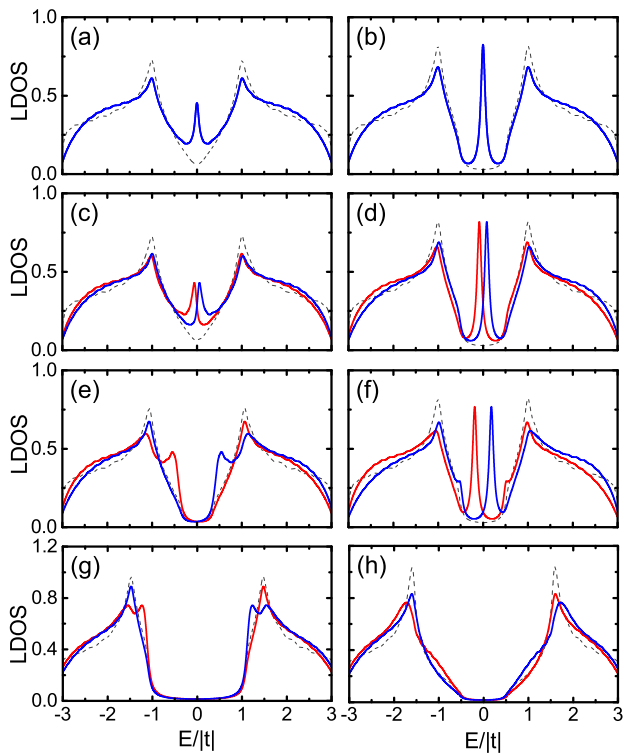}
\caption{(color online). LDOS for $\lambda=0.0$ [left column,
including (a),(c),(e),(g)] and for $\lambda=0.1t$
 [right column, including (b),(d),(f),(h)], where the Hubbard interaction $U=0.0$ for (a) and (b),
 $U=1.6t$ for (c) and (d), $U=2.6t$ for (e) and (f), and $U=3.6t$ for (g) and (h), respectively. LDOS for different spins is resolved, those with
 the spin up are denoted by the blue lines and the spin down the red lines. The grey dash lines represent the LDOS on the lattice site away from the vacancy. }\label{loc2}
\end{center}
\end{figure}

In order to show the emergence of the magnetism induced by the
vacancy in more detail, we calculate the spin resolved local
density of state(LDOS) as defined by,
\begin{equation}
D_{\sigma}(\epsilon)=\Sigma_{n,i}|u^{n}_{i,\sigma}|^2\delta(\epsilon-\epsilon_n),
\end{equation}
where $i$ runs over the lattice sites surrounding the vacancy up
to the third-nearest neighbors, as those linked by the green line
in Fig.\ref{loc1}(a) and (b). $u^{n}_{i,\sigma}$ is the
single-particle amplitude on the $i$th site with spin $\sigma$ and
the corresponding eigenvalue is $\epsilon_n$. The Delta function in Eq.(2) is
replaced by the Lorentzian function for plotting. The results for the LDOS are presented
in Fig.~\ref{loc2}(a)-(h) for different Hubbard interaction $U$ and SO interaction
$\lambda$. The red and blue lines
represent the LDOS for the spin up and spin down components
respectively, and the dash lines show the LDOS away from the
vacancy for a comparison. In the case of $U=\lambda=0.0$ as shown in
Fig.~\ref{loc2}(a), the LDOS shows a V-shape linear behavior near
the Fermi level for those lattice sites far away from the vacancy
(denoted by the dashed line) which is the consequence of the
linear dispersion relation of the electrons, the so-called Dirac
fermions. For those around the vacancy, a peak at the Fermi level
emerges as shown by the solid line, which corresponds to the
localized states induced by the vacancy~\cite{Castro}. After turning on the SO
coupling, such as that for $\lambda=0.1t$[see Fig.2(b)], we can
see that an energy gap opens for those lattice sites far away
from the vacancy~\cite{Kane,Kane2}, so that now a U-shape LDOS near the Fermi level occurs. In this way, the mid-gap peak is enhanced
noticeably because the decay rate of the localized states into
the continuum is reduced largely due to the open of the energy
gap. This will lead to the increase in the spectral weight of the
localized states around the vacancy. However, for both cases, one will
find that the LDOS for the spin up and spin down components degenerates, so that
the system will not show magnetism as
a whole without the Hubbard interaction.

The effect of a finite Hubbard interaction $U$ is to split the
spin degenerate LDOS, so that two peaks occur corresponding to different
spins, as shown in Fig.~\ref{loc2}(c)-(f).
Consequently, the localized spin up and down moments will not cancel out in this case, and a net magnetism around the vacancy is induced.

\begin{figure}[pt]
\begin{center}
\includegraphics[width=0.44\textwidth]{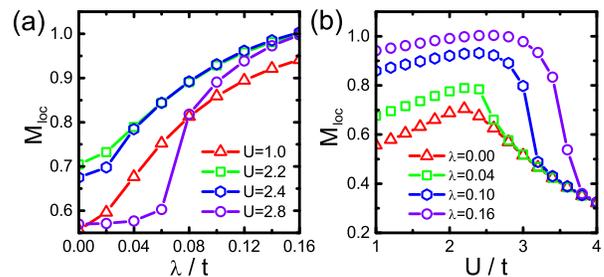}
\caption{(color online). Local moments $M_{loc}$ (see text) are shown as a function of the SO coupling
$\lambda$ for different Hubbard interaction $U$(a) and of $U$ for different $\lambda$(b).}\label{loc3}
\end{center}
\end{figure}

The magnetism may be quantified by the local moment
$M_{loc}=\sum_{i}m_{i}$, where the sum runs over the lattice sites surrounding the vacancy up to the third-nearest neighbors as used above in the calculation for the LDOS. The results are presented in Fig.~\ref{loc3}(a) and (b) for different $U$ and $\lambda$, respectively. The local moment $M_{loc}$ shows a monotonic increase with the SO coupling $\lambda$, so it reinforces our observation that the local magnetism is enhanced by the SO coupling as shown in Fig.1. On the other hand, $M_{loc}$ shows a nonmonotonic dependence on the Hubbard interaction
$U$, namely it increases with $U$ firstly and then decreases with a further increase of $U$ after a critical value $U_{c}$.
As discussed above, the local magnetism is determined by the spin-split localized states induced by the vacancy, and it is the Hubbard interaction $U$ to split the spin-degenerate states. Because the open of the gap due to the SO coupling will decrease the decay rate of the localized states into the continuum, so it will enhance the spectral weight of the localized states[see also Fig.~\ref{loc2}], consequently the localized magnetism. The splitting between the two localized states with different spins is proportional to $U$, so the two split localized states will situate in the SO gap for a small $U$[Fig.~\ref{loc2}(c)-(f)]. However, when $U>U_{c}$ the splitting will be larger than the SO gap, and it pushes the localized states to merge into the continuum[Fig.2(g) and (h)], so the local magnetism will decrease.

\section{THE CASE OF MULTI-SITE VACANCY}

\begin{figure}[b]
\begin{center}
\includegraphics[width=0.46\textwidth]{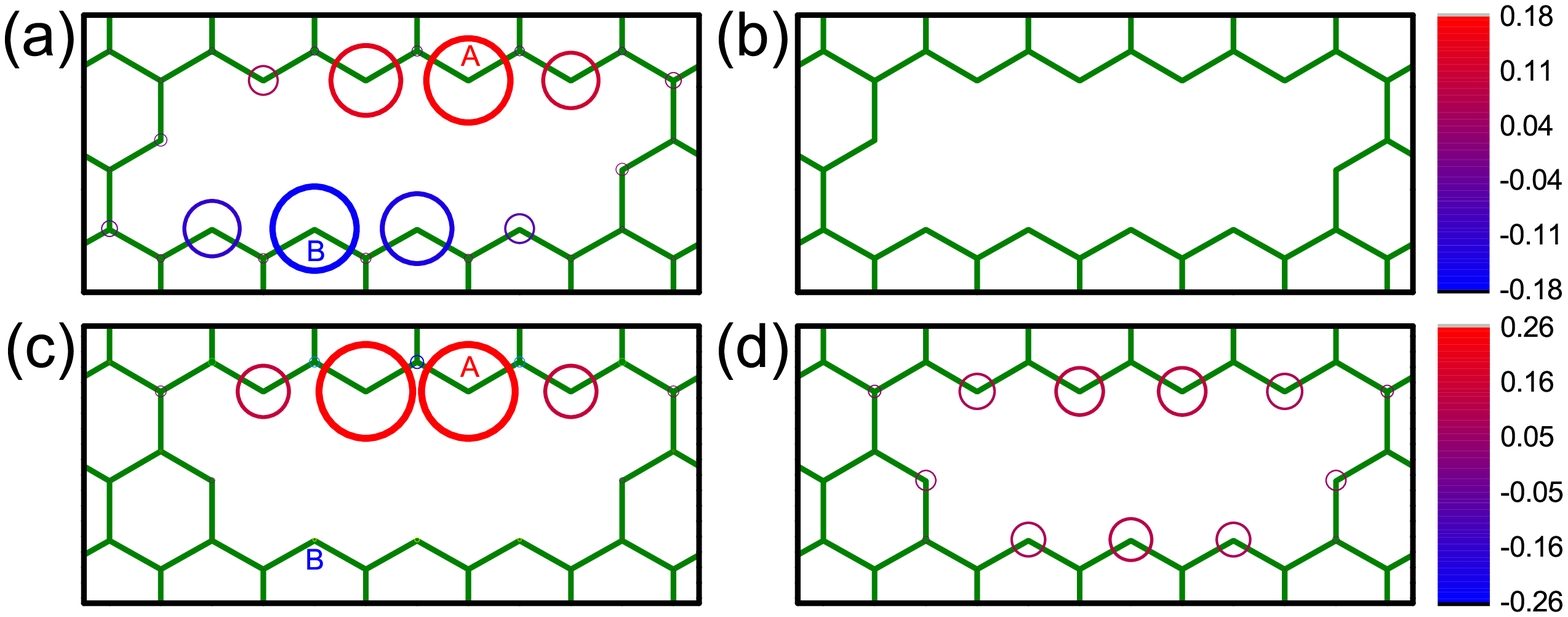}
\caption{(color online).
Distribution of the spin moments $m_{i}$ on the lattice sites surrounding the vacancies for $U=1.0t$ and $L=14$. A cluster of vacancies is formed with the number of missing sites for (a), (b) $N_{v}=8$ and (c), (d) $N_{v}=7$. SO coupling is set to be $\lambda=0.0$ for (a), (c) and $\lambda=0.1t$ for (b), (d). The area and color of hollow circles represent the magnitude of the moments. }\label{loc4}
\end{center}
\end{figure}

The multi-site vacancy can be formed by removing the sites continuously. Here, we consider a large stripe vacancy by taking out a chain of sites from the lattice as illuminated in Fig.~\ref{loc4}. In this way, the stripe vacancy consists of one upper and one lower zigzag edges. As clarified by the Lieb theorem~\cite{Lieb}, the sublattice imbalance between the number of atoms belonging to different sublattices will have significant effect on the magnetism. For the stripe vacancy considered here, the imbalance is expressed by the parity of the number of vacancies, where the number is even ($N_{A}=N_{B}$) in Fig.~\ref{loc4}(a) and (b), and odd ($N_{A}\neq N_{B}$) in Fig.~\ref{loc4}(c) and (d), thus the total spin of the system is $S=0$ and $1/2$ respectively.

In the case of even number of vacancies, a ferrimagnetic spin order emerges on both the upper and lower zigzag edges around the stripe vacancies when there is no SO coupling, as shown in Fig.~\ref{loc4}(a). The ferrimagnetic arrangement and the magnitude of the spin moments on these two edges are symmetric, but they are counter-polarized, so they cancel out exactly and the whole system will not show magnetism. This is consistent with the Lieb theorem~\cite{Lieb}.
The ferrimagnetic order on a sufficiently long zigzag edge around the stripe vacancies here is similar to the spin order formed at the outer edge of the zigzag ribbon~\cite{Fujita,Wakabayashi,Young,Fern3} and the graphene nanoisland~\cite{Pala}.
In the case of odd number of vacancies, a similar ferrimagnetic spin order is also induced with a slightly large magnitude [Fig.~\ref{loc4}(c)]. Interestingly, this ferrimagnetic order occurs only on the upper zigzag edge, not on the lower edge. This phenomenon is ascribed to the presence of an extra spin when a sublattice imbalance $N_{A}\neq N_{B}$ exists, as described by the Lieb theorem~\cite{Lieb}.

\begin{figure}[pt]
\begin{center}
\includegraphics[width=0.48\textwidth]{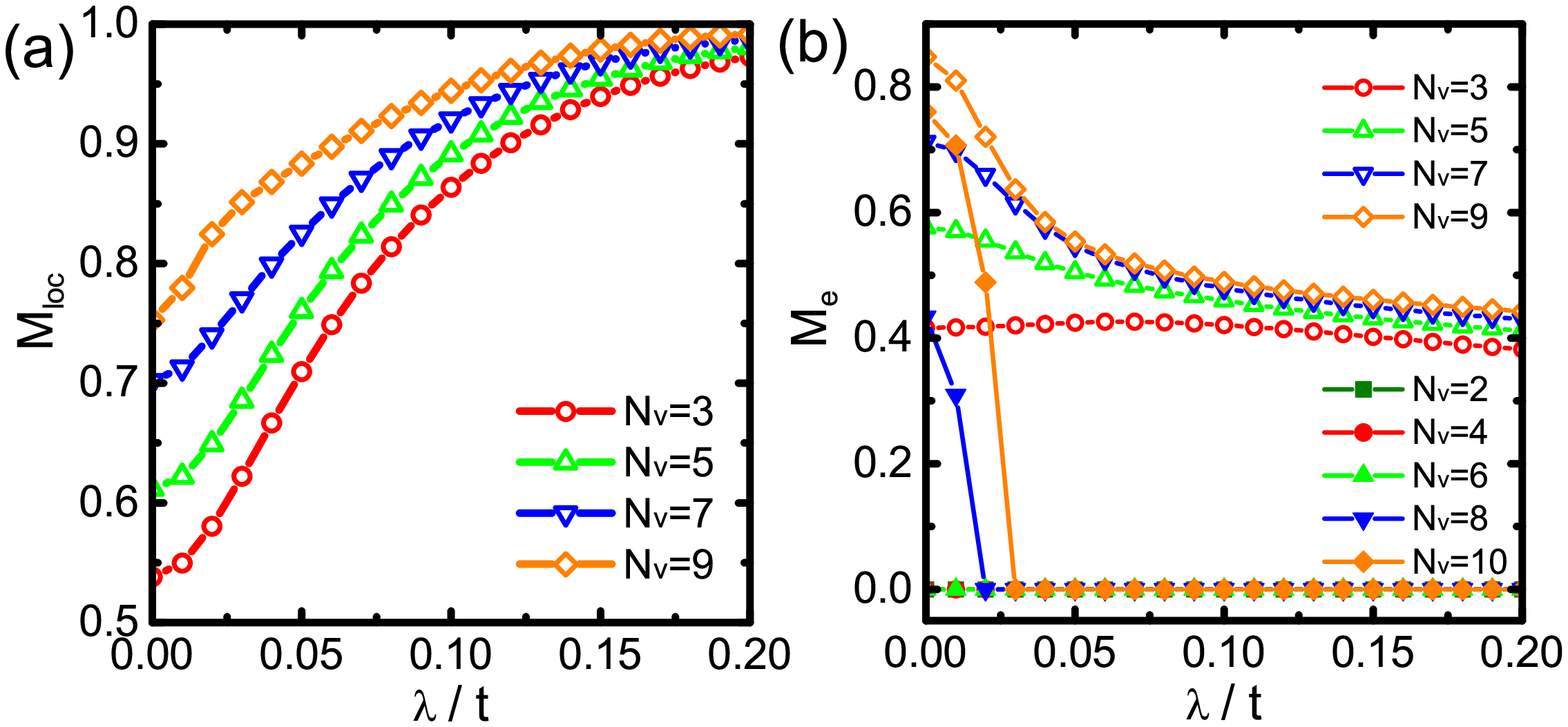}
\caption{(color online). (a)Local moments $M_{loc}$ are plotted as a function of $\lambda$ while $U=1.0t$ and $L=14$. The function in different size of vacancy is distinguished by different color and shape of points. The cases of even $N_{v}$ are not plotted as local moments are always zero obeying Lieb theorem~\cite{Lieb}. (b) The function of edge moments $M_{e}$ versus $\lambda$ are given in different $N_{v}$. }\label{loc5}
\end{center}
\end{figure}

After turning on the SO coupling, such as for $\lambda=0.1t$, the ferrimagnetic spin order on both the upper and lower zigzag edges around the stripe vacancies disappears completely in the case of even number of vacancies[Fig.~\ref{loc4}(b)]. However, the effect of the SO coupling on local magnetism is quite different for the case of an odd number of vacancies. Here, a ferrimagnetic spin order similar to that on the upper edge emerges on the lower edge, though the magnitude of the individual spin moment is reduced[Fig.~\ref{loc4}(d)].
To show variation of the total magnetism, we plot the quantity $M_{loc}$ as a function of the SO coupling $\lambda$ in Fig.~\ref{loc5}(a), here $M_{loc}$ is the sum of the spin moments on the sites which are on the zigzag edges around the vacancies. Since $M_{loc}$ is always zero in the case of even $N_{v}$, it is not plotted here. With an odd $N_{v}$, the local moment $M_{loc}$ increases with the increase of $\lambda$, which shows a similar behavior as that in the case of a single vacancy. This indicates that the total local magnetism shown in Fig.~\ref{loc4}(d) is in fact enhanced with the introduction of the SO coupling and approaches the saturation value 1 finally. From Fig.~\ref{loc5}(a), one can also find that $M_{loc}$ increases with the increase of the number of vacancies $N_{v}$. This suggests that the SO coupling will localize the induced spin moments to those lattice sites which are neighboring the vacancies.

To quantify the variation of the spin moments with $\lambda$ on the upper zigzag edge, we also present $M_{e}$ as a function of  $\lambda$ in Fig.~\ref{loc5}(b), here $M_{e}$ is the sum of the spin moments only on the sites on the upper zigzag edge.
Let us consider firstly the case of even number of $N_{v}$, for a small number of even vacancies, $M_{e}$ is always zero. Up to $N_{v}\geq8$, a finite $M_{e}$ occurs and it increases with $N_{v}$ by the formation of the zigzag edges.
However, $M_{e}$ drops rapidly to zero after turning on the SO coupling. These results quantify the physical picture derived from Fig.~\ref{loc4}(a) and (b).
Now let us turn to the case of odd number of $N_{v}$. Without the SO coupling, $M_{e}$ also shows an increase with $N_{v}$. With the introduction of the SO coupling, $M_{e}$ shows a decrease with $\lambda$ and saturates to near one half of ${M_{loc}}$.

\begin{figure}
\begin{center}
\includegraphics[width=0.48\textwidth]{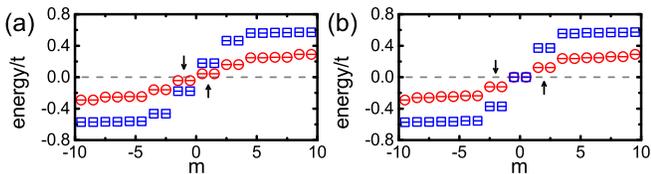}
\caption{(color online). The single-particle energy levels labeled with $m$ (see text) near the Fermi level of the non-interacting systems for (a)$N_v=8$ and (b) $N_v=7$. SO coupling is set to be $\lambda=0.0$ for the red circles and $\lambda=0.1t$ for the blue squares.}\label{loc6}
\end{center}
\end{figure}

In fact, we can make an analogy between the stripe vacancy and the graphene ribbon with zigzag edges. A remarkable feature of the graphene ribbon with zigzag edges is that it has a flat band localized on the zigzag edge\cite{Neto}. An important effect of this flat band is that a counter-polarized ferromagnetic order along the upper and lower edges will be induced when the Hubbard interaction between electrons is included ~\cite{Fujita,Wakabayashi,Young,Fern3}. In view of this, we plot the single-particle spectra for the systems with the stripe vacancy without the Hubbard interaction in Fig.\ref{loc6}(a) and (b) for $N_{\nu}=8$ and $N_{\nu}=7$, respectively. Each energy level is labeled with $m=n-N_{e}-1/2$ in order to indicate that the energy level with $m<0$ is occupied by electron. For the systems without SO coupling, we find that there are four near-degeneracy localized states [red circles indicated by arrows in Fig.\ref{loc6}(a) and (b)] which is near the Fermi level for both $N_{\nu}=8$ and $N_{\nu}=7$. These states will have the same effect as the flat band in the zigzag ribbon when a suitable Hubbard $U$ is turned on. So, a counter-polarized ferrimagnetic order as shown in Fig.\ref{loc4}(a) will emerge. However, we note that there are two additional zero modes for $N_{\nu}=7$ relative to $N_{\nu}=8$, due to the imbalance between the sublattices ($N_{A}>N_{B}$). These zero modes will induce extra spin moments on both edges, which counteract the antiparallel moments on the lower edge. Thus, in the case of $N_{\nu}=7$, only the ferrimagnetic order on the upper edge appears.
After turning on the SO coupling, those localized states [blue squares indicated by arrows in Fig.\ref{loc6}(a) and (b)] are pushed away from the Fermi level due to the open of the SO gap. Thus, as shown in Fig.\ref{loc4}(b), a small Hubbard U is not enough to induce the counter-polarized ferrimagnetic order on the upper and lower edges. However, the zero modes originating from the imbalance of sublattices are not affected by the SO coupling [Fig.\ref{loc6}(b)]. So, the additional ferrimagnetic order on both edges induced by these zero modes will remain for $N_{\nu}=7$.

\section{CONCLUSION}

In a summary, we have studied the local magnetism induced by vacancies on the honeycomb lattice based on the Kane-Mele-Hubbard model.
It is shown that the SO coupling tends to localize and consequently enhances the local magnetic moments near a single vacancy. Furthermore, along the zigzag edges formed by a chain of vacancies, the SO coupling will suppress completely the counter-polarized ferrimagnetic order along the edges. Therefore, the system will not show any local magnetism in the case of even number of vacancies. For an odd number of vacancies, a ferrimagnetic order along both edges exists and the total magnetic moments along both edges will increase.

\begin{acknowledgments}
This work was supported by the National Natural Science Foundation
of China (Grant Nos. 91021001, 11190023 and 11204125) and the Ministry of Science and
Technology of China (973 Project grant numbers 2011CB922101 and
2011CB605902).
\end{acknowledgments}

\end{document}